\documentclass[journal, draftclsnofoot, onecolumn, 12pt]{IEEEtran}

\usepackage{blindtext}
\usepackage{caption}
\usepackage{subcaption}
\usepackage{graphicx}
\usepackage{setspace}
\usepackage{epstopdf}
\usepackage{amsfonts}
\usepackage{newtxtext}
\usepackage{amsmath}
\usepackage{balance}
\usepackage{amssymb}
\usepackage[latin9]{inputenc}
\usepackage{centernot}
\usepackage{bm}
\usepackage{stmaryrd}
\usepackage{float}
\usepackage{array}
\usepackage[noadjust]{cite}
\usepackage[thmmarks]{ntheorem}
\usepackage{pdfpages}
\usepackage{bbold}
\usepackage{centernot}
\usepackage{csquotes}
\usepackage{algorithm}
\usepackage{algpseudocode}
\usepackage{mathtools}
\usepackage{setspace}
\usepackage{tikz}
\usepackage[margin=1in]{geometry}
\usepackage{comment}
\usetikzlibrary{calc,matrix,positioning}
\tikzset{ 
table/.style={
  matrix of nodes,
  row sep=-\pgflinewidth,
  column sep=-\pgflinewidth,
  nodes={rectangle,text width=3em,align=center},
  text depth=1.25ex,
  text height=2.5ex,
  nodes in empty cells
},
}

\makeatletter
\algnewcommand{\LineComment}[1]{\Statex \hskip\ALG@thistlm \(\triangleright\) #1}
\makeatother
\spacing{1.5}

\newtheorem{proposition}{Proposition}

\theoremheaderfont{\bfseries}
\theorembodyfont{\normalfont}
\theoremseparator{:}
\newtheorem{definition}{Definition}
\theoremsymbol{$\blacksquare$}

\newlength\myindent
\newcommand{\lt}{<}
\newcommand\myeq{\mathrel{\overset{\makebox[0pt]{\mbox{\normalfont\tiny\sffamily def}}}{=}}}

\begin{document}

\title{On the Base Station Association Problem in HetSNets}

\author{Zoubeir~Mlika,~\IEEEmembership{Student~Member,~IEEE,}
        Elmahdi~Driouch,~\IEEEmembership{Student~Member,~IEEE,}
        Wessam~Ajib,~\IEEEmembership{Member,~IEEE,}
        and~Halima~Elbiaze,~\IEEEmembership{Member,~IEEE}\\
        
        \IEEEauthorblockA{Department of Computer Science,
        UQAM University\\
        Email: mlika.zoubeir@courrier.uqam.ca,
        driouch.elmahdi@uqam.ca,
        ajib.wessam@uqam.ca,
        elbiaze.halima@uqam.ca}
}

\maketitle

\begin{abstract}
The dense deployment of small-cell base stations in HetSNets requires efficient resource allocation techniques. More precisely, the problem of associating users to SBSs must be revised and carefully studied. This problem is NP-hard and requires solving an integer optimization problem. In order to efficiently solve this problem, we model it using non-cooperative game theory. First, we design two non-cooperative games to solve the problem and show the existence of pure Nash equilibria (PNE) in both games. These equilibria are shown to be far from the social optimum. Hence, we propose a better game design in order to approach this optimum. This new game is proved to have no PNE in general. However, simulations show, for Rayleigh fading channels, that a PNE always exists for all instances of the game. In addition, we show that its prices of anarchy and stability are close to one. We propose a best response dynamics (BRD) algorithm that converges to a PNE when it exists. Because of the high information exchange of BRD, a completely distributed algorithm, based on the theory of learning, is proposed. Simulations show that this algorithm has tight-to-optimal performance and further it converges to a PNE (when existing) with high probability.\let\thefootnote\relax\footnotetext{This work has been partly presented at IEEE ICC, London, UK, June 2015.}
\end{abstract}


\IEEEpeerreviewmaketitle

\section{Introduction}

The unprecedented growth of mobile data traffic gives rise to many challenges in today's cellular networks. Hence, novel network architectures and resource allocation solutions are needed in order to deal with this exponential growth. Cellular networks have shifted from the deployment of traditional and expensive high-power base stations (BSs) towards the deployment of heterogeneous low-power BSs including small-cell BSs (SBSs)~\cite{Ghosh}. This led to the appearance of heterogeneous and small-cell networks (HetSNets). In HetSNets, different heterogeneous elements coexist such as distributed antenna systems, relay nodes, and SBSs including pico-cell BSs (PBSs), femto-cell BSs (FBSs), etc. The deployment of SBSs is emerging as a key solution to the increasing demand for reliable and high speed wireless access~\cite{AndrewsJ}. Their adoption is motivated by several factors such as their ease of deployment and low cost of installation and maintenance. They are regarded as the ideal candidate for future generation of cellular network in order to enable better coverage and achieve higher data rates~\cite{Hossain}.

In HetSNets, two kinds of interference arise, namely cross-tier interference and co-tier interference~\cite{Saquib}. Whereas the former kind can be managed using the spectrum splitting approach~\cite{Chandrasekhar}, where macro-cell BSs (MBSs) and SBSs are allocated different portions of the spectrum, the co-tier interference is very difficult to control especially when there is no communication between the deployed SBSs. To better manage the co-tier interference, several techniques are introduced in the literature such as power allocation, spectrum allocation, and user-BS association~\cite{AndrewsJ}. In this paper, we are interested in the user-BS association problem. Roughly speaking, we define this problem as follows: given a set of users, a set of BSs, a threshold value and a channel gain between every pair of user-BS, find a one-to-one association of the users to the BSs such that every association has a signal-to-interference-plus-noise ratio (SINR) greater than the threshold value. In our previous work, we formulated this problem and proved that it is NP-hard~\cite{Zoubeir}. In this paper, we are interested in solving it distributively in HetSNets. Hence, we propose to model it using non-cooperative game theory. Moreover, we design two strategy updating algorithms which approach the highly complex centralized user-BS association. The variety of the proposed techniques in the literature either do not use game theory or do not propose a completely distributed algorithms for such problem. This motivates the design of efficient game models and completely distributed user-BS association mechanisms in HetSNets.

Related work can be divided into: (\textit{i}) centralized solutions where the decision lies on a central coordinator~\cite{Kuang, Li, RuoyuS, YichengL}; (\textit{ii}) distributed solutions with high amount of information exchange~\cite{MingyiH,Qiaoyang, KaimingS}; and (\textit{iii}) applications of learning algorithms for HetSNets. In the sequel, we present the most recent related work on these directions. 

%Some of the related work that solve the problem centrally are the following. 
In~\cite{Kuang}, the authors study the resource allocation as a joint optimization problem of channel allocation, user-BS association, beam-forming and power control in HetSNets. It is solved using an iterative algorithm based on $\ell^1$-norm heuristics. This work maximizes the total up-link throughput and guarantees the QoS of users. Even though, the work shows that the relaxation of the combinatorial problem to a continuous one provides the optimal solution, the proof lacks of generality and it depends on the problem formulation. In~\cite{Li}, the joint power allocation and user-BS association is modeled as a combinatorial optimization problem. The authors use Bender's decomposition to solve the modeled problem optimally and they propose heuristic algorithms. However, the proposed optimal method and the heuristic algorithms are highly complex. In~\cite{RuoyuS}, the joint user-BS association and power control problem in HetSNets is considered. The joint problem is formulated as an optimization problem where the objective is to maximize the minimum SINR subject to a power constraint at each BS. This problem is shown to be NP-hard and heuristic algorithms are proposed to solve it. Note that the differences with our work are threefold. First, the paper solves the user-BS association and the power control in a centralized fashion. Second, the system model allows multiple users to be associated to one BS. Third, the problem does not guarantee a minimum quality of service (QoS) to the associated users. Reference~\cite{YichengL} considers the problem of user-BS association and spectrum allocation in HetSNets. The paper adopts stochastic geometry to derive the theoretical mean utility based on the coverage rate. The user-BS association is performed based on the biased received power and the bias factors are obtained analytically using the cell range expansion scheme. 

%Some of the related works that tackle the problem distributively are the following. 
In~\cite{MingyiH}, the authors solves the joint problem of user-BS association and resource allocation in orthogonal-frequency-division-multiple-access (OFDMA) networks. The problem is formulated as a weighted sum rate maximization problem and is shown to be NP-hard. Next, based on mechanism design, the joint problem of user-BS association and resource allocation is modeled using non-cooperative game and solved using Vickrey-Clarke-Groves (VCG) mechanism. The main differences between this work and ours are twofold. First, the optimization objective is a weighted sum rate which does not depend on the interference of the associated users, i.e., because of the OFDMA system model, there is no interference and therefore the rate is a function of the signal-to-noise ratio (SNR). Second, the user-BS association game has a continuous utility function (the weighted rate of the user) which allows the authors to use the theory of super-modularity and complementarity whereas we consider a discontinuous utility function and different system model. The user-BS association problem is solved in~\cite{Qiaoyang} jointly for fairness and load balancing. The load of the SBSs is balanced by using a distributed algorithm based on the technique of dual decomposition. This work solves the user-BS association problem based on relaxation and rounding techniques which remove the combinatorial nature of the problem and render it easier to solve. Reference~\cite{KaimingS} solves the user-BS association in HetSNets based on a pricing scheme. The user-BS association is solved based on a Lagrangian dual analysis and a dual coordinate descent method is proposed. The paper also extends the problem to the multiple-input-multiple-output (MIMO) case and optimizes the beam-forming variables. Anyhow, the proposed distributed algorithm is not based on game theory. Both~\cite{Qiaoyang} and~\cite{KaimingS} focus on the same system model which is very different from the system model of this paper.

The related work that investigate the application of learning algorithms in HetSNets focus on other types of problems such as power allocation and link activation~\cite{Andrews, Dinitz, Rose, Nazir}. In~\cite{Andrews}, the authors formulate the link activation problem as a non-cooperative game and study its convergence to a mixed NE. The transmitter-receiver links are already established whereas the problem is to determine which links are to be simultaneously activated. The same system model is considered in~\cite{Dinitz} where the author designed no-regret algorithms that converge to sub-optimal performance. Note that once again, the tackled problem in this work is power allocation and the links are pre-established. The work in~\cite{Rose} makes use of a recent learning paradigm called interactive trial and error in order to design completely distributed algorithms for the power allocation problem. The authors prove the convergence of their solution to an epsilon-NE. In the context of femto-cell networks, many learning algorithms were proposed in the literature. For instance, the authors in~\cite{Nazir} propose and compare two learning mechanisms based on $Q$-learning and evolutionary game theory.

The key contributions of this paper are the following:
\begin{itemize}
	\item We model the user-BS association problem in a HetSNet as non-cooperative repeated games.
	\item We investigate the pure Nash equilibrium (PNE) as a solution concept for the formulated games and we prove, using potential game and best response update, that these games admit PNEs.
	\item We show, using simulations, that the proposed games have low performance compared to the social optimum.
	\item To overcome these shortcomings, we propose a better game design. This new game is shown to have no PNE in general however. Despite this, when the reformulated game admits PNEs, these PNEs are very efficient. In other words, the price of stability (PoS) and the price of anarchy (PoA) of the game are shown to be very close to one. Furthermore, using simulations, it is observed that the game always admits PNEs for Rayleigh fading channels. 
	\item In order to solve the formulated game, we propose two strategy updating algorithms. The first is the best response dynamics (BRD) algorithm. The second is a completely distributed algorithm, inspired by the well-known learning rule \textit{win-stay-lose-shift} (WSLS)~\cite{Nowak, Mihaylov} and is called \textit{modified} mWSLS.
	\item We show that mWSLS can be implemented in a completely distributed manner and that it has tight-to-optimal performance as well. Furthermore, the efficiency of mWSLS is shown to converge to a PNE (when it exists) with high probability.
\end{itemize}
%The user-BS association problem is formulated as integer linear programming problem and solved using CPLEX branch-and-bound solver~\cite{CPLEX}. This centralized optimal solution is used as a benchmark solution to compare other proposed algorithms against.

The rest of the paper is organized as follows: Section~\ref{sys} presents the system model and formulates the integer programming problem. Section~\ref{game} first formulates the user-BS association problem as two non-cooperative games and shows the existence of PNEs in both games. Second, it designs a better game and analyzes it. Section~\ref{disc} discusses the performance and the efficiency of the PNEs. Next, Section~\ref{BRD} describes the proposed best response dynamics (BRD) algorithm and Section~\ref{mwsls} proposes a completely distributed algorithm. Then, Section~\ref{simu} presents the simulation results and finally Section~\ref{cl} draws some conclusions.

\section{System Model\label{sys}}

This paper considers a HetSNet composed of a macro-cell base station (MBS), $N$ SBSs and $M$ small-cell users (SUs). The set of SUs and the set of SBSs are denoted by $\mathcal{M}\myeq\{1, \dotsc, M\}$ and $\mathcal{N}\myeq\{1, \dotsc, N\}$, respectively. The SBSs are assumed to have no cognitive capabilities and hence cannot perform spectrum sensing in order to avoid interference to and from the MBS transmissions. Therefore, the spectrum splitting approach~\cite{Chandrasekhar}, where the SBSs are transmitting over a portion of the MBS spectrum whereas the remaining part of the spectrum is exclusively dedicated to the macro-cell users (MUs), is assumed. Thus, there is no cross-tier interference between the MBS and the SBS transmissions. However, SBSs transmitting at the same time will suffer from co-tier interference. The down-link transmission is considered where each available SBS can transmit to a SU. Every transceiver in the network is equipped with a single antenna and one SBS is transmitting to one and only one SU at any given time. An example of the system model is given in Fig.~\ref{fig:sys}.
\iftrue
\begin{figure}[htbp]
	\centering
	\includegraphics[scale=0.55]{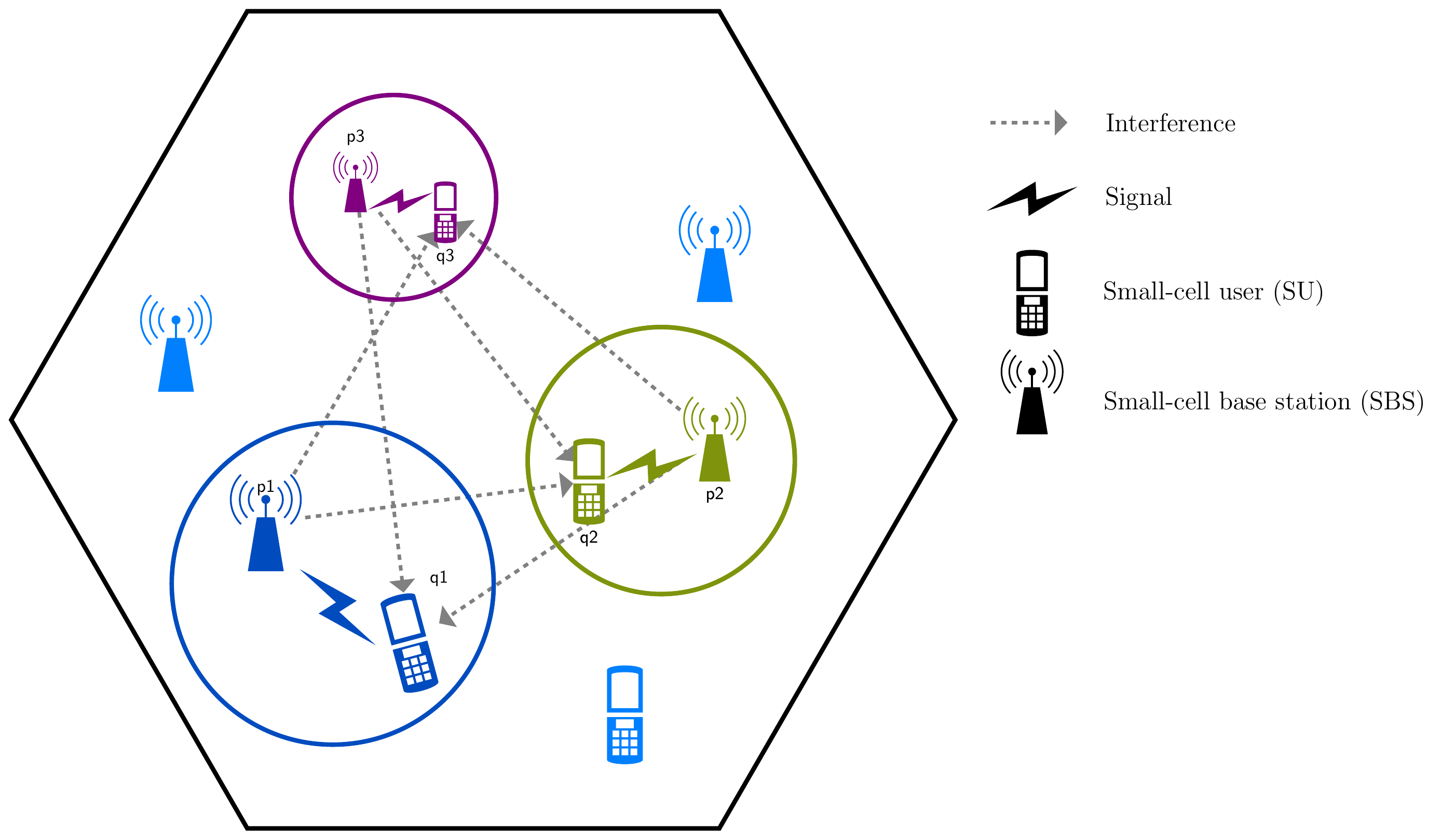}
	\caption{System model.}
	\label{fig:sys}
\end{figure}
\fi

When a SBS $n\in\mathcal{N}$ is transmitting to a SU $m\in\mathcal{M}$, it uses a fixed power $p_{n}$ and it establishes a link $\ell_{mn}$. The channel gain over link $\ell_{mn}$, for all $n\in\mathcal{N}$ and for all $m\in\mathcal{M}$, is modeled as a random variable $\mathrm{g}_{mn}$ which takes into account the short-term path loss propagation effect and the long-term fading effect. Further, time is divided into time-slots and the channel gains are fixed during one time-slot. The received SINR of the established link $\ell_{mn}$ is given by:

\begin{align}
\label{sinr:1}
\mathrm{SINR}_{mn}\myeq \frac{p_{n}\cdot|\mathrm{g}_{mn}|^2}{\sigma_n^2 + \sum\limits_{n'\in\mathcal{F}'}p_{n'}\cdot|\mathrm{g}_{mn'}|^2},
\end{align}
where $\mathcal{F}'=\mathcal{F}\setminus\{n\}$, $\mathcal{F}$ is the set of transmitting SBSs during the current time-slot, and $\sigma_n^2$ is the variance of the additive white Gaussian noise (AWGN). 

In this work, the main objective is to maximize the number of associated SUs in each time-slot. This objective has to be met subject to the constraint of satisfying a system level performance, namely the QoS of associated SUs which has to be kept above a specified threshold. The QoS constraint is expressed using the SINR of each established link. The optimization problem is formulated in the next subsection.

\subsection{Centralized Solution}

The user-BS association problem can be formulated as an integer programming problem~\cite{Zoubeir}. The objective is to maximize the total number of associated SUs in the HetSNet subject to the SINR thresholds constraints at the SUs.

The problem can be formulated as follows:

\begin{subequations}
\label{unw:pb}
\begin{align}
\underset{\mathbf{x}}{\text{maximize}}
        & \quad\sum_{\substack{m\in\mathcal{M}}}\sum_{\substack{n\in\mathcal{N}}}x_{mn}\label{eq:cost}\\
    \text{subject to} 
        & \quad\sum_{m\in\mathcal{M}}x_{mn}\leqslant1,\;\forall\;n\in\mathcal{N},\label{cns:1}\\
        & \quad\sum_{n\in\mathcal{N}}x_{mn}\leqslant1,\;\forall\;m\in\mathcal{M}, \label{cns:2}\\
        & \quad\Gamma_{mn}\left(\mathbf{x}\right)\geqslant\beta\cdot x_{mn},\;\forall\;m\in\mathcal{M},\;\forall\;n\in\mathcal{N},\label{cns:3}\\
        & \quad x_{mn}\in\left\{{0, 1}\right\},\;\forall\;m\in\mathcal{M},\;\forall\;n\in\mathcal{N}. \label{cns:4}
\end{align}
\end{subequations}

The optimization variables are given by the matrix $\mathbf{x}=\left[x_{mn}\right]$. Constraints \eqref{cns:1} ensure that a SBS associates to one SU whereas constraints \eqref{cns:2} ensure that a SU is associated with one SBS. Constraints \eqref{cns:3} guarantee that an associated SU-SBS must have an SINR above the threshold $\beta$. Finally, constraints \eqref{cns:4} ensures that association variables $x_{mn}$ are Boolean. Note that \eqref{cns:3} represents the SINR of the corresponding link $\ell_{mn}$ and is given by:
\begin{align}\label{sinr:mn}
\Gamma_{mn}\left(\mathbf{x}\right)\myeq\frac{p_n\cdot|\mathrm{g}_{mn}|^2\cdot x_{mn}}{\sigma_n^2+\sum\limits_{\substack{m^\prime\in\mathcal{M}^\prime}}\sum\limits_{\substack{n^\prime\in\mathcal{N}^\prime}}p_{n'}\cdot |\mathrm{g}_{mn^\prime}|^2\cdot x_{m^\prime n^\prime}},
\end{align}
where $\mathcal{M}^\prime=\mathcal{M}\setminus\{m\}$,  $\mathcal{N}^\prime=\mathcal{N}\setminus\{n\}$.

We proved in~\cite{Zoubeir} that problem \eqref{unw:pb} is NP-hard. Therefore, its optimal solution cannot be found in polynomial-time unless $\mathrm{P}=\mathrm{NP}$. In this paper, problem \eqref{unw:pb} is solved using a branch-and-bound algorithm implemented in the CPLEX solver~\cite{CPLEX}. This algorithm gives the optimal solution for reasonable values of $M$ and $N$ and therefore it is used in the subsequent sections as a benchmark solution.

%In the next sections, the formulated problem is first modeled and studied using non-cooperative game theory. Next, a distributed algorithm is proposed to solve the formulated game.

\section{User-BS Association Games\label{game}}
\subsection{Games Formulation}

This section formulates the user-BS association problem using non-cooperative game theory. First, we propose a simple game model which is shown to admit PNEs. Second, we modify this game by restricting the players to not play a collision; that is, none of the players chooses the same action.

The first user-BS association game is given by $\mathfrak{G}_1\myeq\langle\mathcal{N}, \{\mathcal{B}_n\}_{n\in\mathcal{N}}, \{v_{n}\}_{n\in\mathcal{N}}\rangle$ where:

\begin{itemize}
    \item $\mathcal{N}$ is the set of players, i.e., the SBSs\footnote{We use SBSs (resp. SUs) and players (resp. users) interchangeably.};
    \item $\{\mathcal{B}_n\}_{n\in\mathcal{N}}$ is the set of actions available for player $n$ and is given by $\mathcal{B}_n = \mathcal{M}$.
    The actions of all the players is given by the Cartesian product $\mathcal{B}=\mathcal{B}_1\times\mathcal{B}_2\times\cdots\times\mathcal{B}_N$. The vector $\mathbf{a} = (\mathrm{a}_1, \dotsc, \mathrm{a}_N) \in \mathcal{B}$ denotes an action profile of $\mathfrak{G}_1$, where $\mathrm{a}_n\in\mathcal{B}_n$; and
    \item $\{v_{n}\}_{n\in\mathcal{N}}$ is the set of payoffs of the players. The payoff of player $n$ is given by the function $v_{n}:\mathcal{B}\mapsto\{-1,1\}$.
\end{itemize}
According to these notations, we rewrite the SINR expression in \eqref{sinr:1} as follows:
\begin{align}
\label{sinr:a}
	\mathrm{SINR}_{\mathrm{a}_nn}(\mathbf{a})=\mathrm{SINR}_{\mathrm{a}_n n}\left(\mathrm{a}_n, \mathbf{a}_{-n}\right)\myeq \frac{p_{n}\cdot|\mathrm{g}_{\mathrm{a}_nn}|^2}{\sigma_n^2 +\sum\limits_{n'\in\mathcal{S}_n}p_{n'}\cdot|\mathrm{g}_{\mathrm{a}_nn'}|^2},
\end{align}
where $\mathcal{S}_n\myeq\{n'\in\mathcal{N}: n'\neq n\}$, $\mathbf{a}$ is an action profile in $\mathcal{B}$, $\mathbf{a}_{-n}$ is the action profile $\mathbf{a}$ where player's $n$ action is dropped, i.e., $\mathbf{a}_{-n}=\left(\mathrm{a}_1,\cdots, \mathrm{a}_{n-1},\mathrm{a}_{n+1},\cdots, \mathrm{a}_{N}\right)$.

The payoff function of player $n$, $v_n(\cdot)$, is given by:
\begin{align}
    \label{cases:3}
    v_n(\mathbf{a}) = v_n(\mathrm{a}_{n}, \mathbf{a_{-n}})\myeq
    \begin{cases}
        -1&\mbox{if }\, \mathrm{SINR}_{\mathrm{a}_n n}\left(\mathrm{a}_n, \mathbf{a}_{-n}\right)\lt\beta\\
        1&\mbox{if }\, \mathrm{SINR}_{\mathrm{a}_n n}\left(\mathrm{a}_n, \mathbf{a}_{-n}\right)\geqslant\beta
    \end{cases}.
\end{align}
The following definitions are used in the rest of the paper.
\begin{definition}[Pure Nash Equilibrium]
	A PNE is an action profile $\mathbf{a}^*=\left(\mathrm{a}_n^*, \mathbf{a}_{-n}^*\right)$ such that for all $\mathrm{a}_n'\mathcal{A}_n$, $u_n\left(\mathrm{a}_n^*, \mathbf{a}_{-n}^*\right)\geqslant u_n\left(\mathrm{a}_n', \mathbf{a}_{-n}^*\right)$.
\end{definition}
\begin{definition}[Potential Game]
\label{PG}
    A game is called a \textit{potential game} if the incentive of all players to change their strategy can be expressed using a single global function called the potential function.
\end{definition}
\begin{definition}[Potential Function]
\label{PF}
    A function $\Phi:\mathcal{A}\mapsto\mathbb{R}$ is an \textit{exact potential function} if for every $n\in\mathcal{N}$ and for every $\mathbf{a}_{-n}\in\mathcal{A}_{-n}$: 
\begin{align*}
    \begin{split}
    	\Delta u_n=u_n(\mathrm{a}_n, \mathbf{a}_{-n})-u_n(\mathrm{b}_n, \mathbf{a}_{-n})&=\Phi(\mathrm{a}_n, \mathbf{a}_{-n})-\Phi(\mathrm{b}_n, \mathbf{a}_{-n})\\&=\Delta\Phi\\&\text{for every}\;\mathrm{a}_n,\mathrm{b}_n\in\mathcal{A}_n.
    \end{split}	
\end{align*}
\end{definition}

\begin{proposition}
\label{prop:1}
    The game $\mathfrak{G}_1$ admits at least one PNE.
\end{proposition}

\begin{IEEEproof}
    We prove Proposition~\ref{prop:1} by showing that $\mathfrak{G}_1$ is a potential game.
    Let $\Phi_1:\mathcal{B}=\times_{n=1}^N\mathcal{B}_n\mapsto\mathbb{Z}$ be the function defined as follows:
    \begin{align}
        \Phi_1(\mathrm{a}_{n}, \mathbf{a}_{-n})\myeq\sum_{i=1}^Nv_i(\mathrm{a}_{n}, \mathbf{a}_{-n}).
    \end{align}
	It is easy to see that:
	\begin{align}
		\begin{split}
			\Delta\Phi_1&=\sum_{i=1}^N v_i(\mathrm{a}_{n}, \mathbf{a}_{-n})-\sum_{i=1}^N v_i(\mathrm{b}_{n}, \mathbf{a}_{-n})\\&=v_n(\mathrm{a}_{n}, \mathbf{a}_{-n})+\sum_{i=1,i\neq n}^N v_{i}(\mathrm{a}_{n}, \mathbf{a}_{-n})-v_n(\mathrm{b}_{n}, \mathbf{a}_{-n})-\sum_{i=1,i\neq n}^N v_{i}(\mathrm{b}_{n}, \mathbf{a}_{-n})\\&= \Delta v_n+\underbrace{\sum_{i=1,i\neq n}^N v_{i}(\mathrm{a}_{n}, \mathbf{a}_{-n})-\sum_{i=1,i\neq n}^N v_{i}(\mathrm{b}_{n}, \mathbf{a}_{-n}).}_{=0}\label{pot:fun:1}
		\end{split}
	\end{align}
	The difference of the two sums in the last line of the right-hand side of \eqref{pot:fun:1} is equal to zero because the change in action of player $n$ from $\mathrm{a}_n$ to $\mathrm{b}_n$ only affects player $n$. In fact, if an arbitrary SBS $n$ changes its action from $\mathrm{a}_n$ to $\mathrm{b}_n$, the only payoff that will change is its own payoff since the interference of SBS $n$ to the other SBSs is already there and changing the chosen action does not affect the value of the interference.
	
	According to Definitions~\ref{PG} and~\ref{PF}, the game $\mathfrak{G}_1$ is a potential game and hence admits at least one PNE~\cite{Monderer}.
	%In order to prove that $\mathfrak{G}_2$ has a PNE, we notice that the silent action $s$ is a strongly dominated strategy. Hence, no player will play $s$ at a PNE. The rest of the proof is similar to the one of $\mathfrak{G}_1$.
\end{IEEEproof}

Note that the formulation of $\mathfrak{G}_1$ lacks an important issue in the system model which is occurrence of the collision, where at least two SBSs choose the same action. In order to take into account this issue, we modify $\mathfrak{G}_1$ to the following game given by
$\mathfrak{G}_2\myeq\langle\mathcal{N}, \{\mathcal{B}_n\}_{n\in\mathcal{N}}, \{w_{n}\}_{n\in\mathcal{N}}\rangle$ where the payoff function of player $n$, $w_n(\cdot)$, is given as follows:
\begin{align}
    \label{cases:4}
    w_n(\mathbf{a}) = w_n(\mathrm{a}_{n}, \mathbf{a_{-n}})\myeq
    \begin{cases}
        -2&\mbox{if }\, \exists\, n'\neq n: \mathrm{a}_n = \mathrm{a}_{n'}\\
        -1&\mbox{if }\, \mathrm{SINR}_{\mathrm{a}_n n}\left(\mathrm{a}_n, \mathbf{a}_{-n}\right)\lt\beta\\
        1&\mbox{if }\, \mathrm{SINR}_{\mathrm{a}_n n}\left(\mathrm{a}_n, \mathbf{a}_{-n}\right)\geqslant\beta
    \end{cases}.
\end{align}
%Note that we attribute a payoff value for the collision which stated mathematically as ``$\exists\,n'\neq n:\mathrm{a}_n=\mathrm{a}_{n'}$''. 
According to this formulation, SBSs will try to stay away  from collisions in order to selfishly improve their performance.

In game theory, a best response is an action profile that produces the most favorable outcome for a player, given other players' actions~\cite{Fudenberg}. In fact, the concept of best response is central to Nash's theorem. In other words, the intersection of the best responses of every player is the set of PNE of a given game. The following definition is useful for the next analysis.

\begin{definition}[Best Response]
\label{BR_def}
	Let $\mathrm{BR}\left(\cdot\right)$ be the set of best responses of player $n$. We have the following result:
	$\mathrm{a}_n^{*}\in\mathrm{BR}\left(\mathbf{a}_{-n}\right) \Leftrightarrow \forall\, \mathrm{a}_n\in \mathcal{A}_n, u_n\left(\mathrm{a}_n^{*}, \mathbf{a}_{-n}\right)\geqslant u_n\left(\mathrm{a}_n, \mathbf{a}_{-n}\right)$.
\end{definition}
Definition~\ref{BR_def} says that the action $\mathrm{a}_n^{*}$ of player $n$ is a better response given the action profile $\mathbf{a}_{-n}$ if there is no other action $\mathrm{a}_n$ for player $n$ that do strictly better than $\mathrm{a}_n^{*}$. When each player $n$ plays a best response, we call this process a best response update.
\begin{proposition}
\label{prop:2}
    The game $\mathfrak{G}_2$ admits at least one PNE.
\end{proposition}

\begin{IEEEproof}
    The proof is to show that the best response update of the SBSs converges to an equilibrium point. Assume that we start the best response from an arbitrary action profile $\mathbf{a}$ such that $w_n(\mathbf{a})\in\{1, -1, -2\}$.
    Two cases can be distinguished:
    \begin{enumerate}
    \item $M\geqslant N$ (at least as much actions as players): In this case, SBSs having a payoff of $-2$ change their actions by best responding in order to get $-1$ or $1$. They succeed by doing so because there are more actions than players. Hence, at the end of the first iteration of the best responses, every SBS transmits to a different user and therefore there is no collision. Next, SBSs having a payoff of $-1$ best respond by looking for a SU which is idle (not chosen by other SBSs). If $M=N$, then there is no idle SU and a one-to-one matching exists and none of the SBSs can deviate. Hence, an equilibrium is reached. If $M>N$, then every SBS has to look for the idle SU and transmit to it if possible and an equilibrium is eventually reached.
    \item $M<N$ (more players than actions): In this case, there are some SBSs which are transmitting to the same users (collision). Since there are less actions than SBSs, then the SBSs which are getting $-2$ cannot strictly improve their performance by simply best responding. Hence, the system is already in an equilibrium.
    \end{enumerate}
    This proves that the best response update leads to an equilibrium point where none of the SBSs has the incentive to deviate. Hence, $\mathfrak{G}_2$ admits at least one PNE.
\end{IEEEproof}

In both games $\mathfrak{G}_1$ and $\mathfrak{G}_2$, we proved that they admit PNEs. In $\mathfrak{G}_1$, all SBSs are transmitting in a PNE where some SBSs are in a collision with others whereas in $\mathfrak{G}_2$ the SBSs cannot transmit in a collision state in a PNE for the case where $M\geqslant N$. This deteriorates the performance of the proposed games $\mathfrak{G}_1$ and $\mathfrak{G}_2$ as it will be shown in Section~\ref{simu}.
\iffalse
it is a reason that the game $\mathfrak{G}$ does not have PNEs. In fact, when some of the SBSs can chose to remain silent, some other SBSs can chose to transmit because they can reach the SINR threshold $\beta$ and hence some SBSs will oscillate. An other important remark is the \textit{colliding state}, where two different SBSs chose the same SU. As one can notice, there is no restriction on this colliding state in both games $\mathfrak{G}_1$ and $\mathfrak{G}_2$. This assumption is problematic for the chosen SU when attempting to decode two different signals. However, this is used as a mathematical assumption. Hence, only one SBS will be chosen to transmit to the SU if the SINR is above $\beta$ of course.
\begin{proposition}
\label{prop:1}
    The proposed games $\mathfrak{G}_1$ and $\mathfrak{G}_2$ have PNEs.
\end{proposition}
\fi

The following two definitions will be used in the subsequent sections. 
\begin{definition}[Social Welfare Function]
\label{SW}
	Let $\mathbf{a}$ be an action profile. The social welfare of $\mathbf{a}$ is the sum of the payoffs of all the players, i.e., $\sum_{n}u_n\left(\mathbf{a}\right)$.
\end{definition}
In other words, the social welfare function is the objective function of the optimization problem \eqref{unw:pb} which is given by \eqref{eq:cost}.

\begin{definition}[Social Optimum]
\label{SO}
	An action profile $\mathbf{a}$ is a social optimum when it maximizes the social welfare function.
\end{definition}
In other words, the social optimum is an optimal solution of the optimization problem \eqref{unw:pb}.

The proofs of both propositions~\ref{prop:1} and~\ref{prop:2} allow us to use the simple learning rule \textit{best response dynamics} (BRD) as it is shown to converge to PNEs. Even though games $\mathfrak{G}_1$ and $\mathfrak{G}_2$ are shown to have PNEs, BRD algorithm exhibits poor performance compared to the social optimum solution (as shown in Section~\ref{simu}). Besides, it requires a huge information exchange between the SBSs since each SBS has to know the actions of all other players. To that end, we propose a better game design by reformulating the previous games. Next, we propose the BRD algorithm and a completely distributed algorithm in order to solve the new game.
%, denoted $\mathfrak{G}$. %The discussion of the efficiency of the PNEs and the convergence of the algorithms is eventually presented in Section~\ref{disc}.

\subsection{A Better Game Design}

In this section, we aim to improve the performance of the previously discussed games. Hence, the objective is to design a game that is more efficient and has a performance that is close to the social optimum. To that end, the user-BS association problem is reformulated as a non-cooperative repeated game where the players are the set of the SBSs, $\mathcal{N}$. The set of actions of a player $n\in\mathcal{N}$, $\mathcal{A}_n$, is the set of SUs plus a \textit{silence action}, i.e., $\mathcal{A}_n\myeq\mathcal{M}\cup\{\mathrm{s}\}$, where $\{s\}$ corresponds to the action of not transmitting. The game can be represented in normal form as follows, $\mathfrak{G}\myeq\langle\mathcal{N}, \{\mathcal{A}_n\}_{n\in\mathcal{N}}, \{u_{n}\}_{n\in\mathcal{N}}\rangle$ where:

\begin{itemize}
\item $\mathcal{N}$ is the set of players, i.e., the SBSs;
\item $\{\mathcal{A}_n\}_{n\in\mathcal{N}}$ is the set of actions available for player $n$ and is given by $\mathcal{A}_n = \mathcal{M} \cup \{s\}$.
Similarly, the actions of all the players is given by the Cartesian product $\mathcal{A}=\mathcal{A}_1\times\mathcal{A}_2\times\cdots\times\mathcal{A}_N$. The vector $\mathbf{a} = (\mathrm{a}_1, \dotsc, \mathrm{a}_N) \in \mathcal{A}$ denotes an action profile of $\mathfrak{G}$, where $\mathrm{a}_n\in\mathcal{A}_n$; and
\item $\{u_{n}\}_{n\in\mathcal{N}}$ is the set of payoffs of the players. The payoff of player $n$ is given by the function $u_{n}:\mathcal{A}\mapsto\{-1,0,1\}$. This payoff is viewed as the gain observed by $n$ when choosing to transmit to $m$ (i.e., $\mathrm{a}_n=m$) or when remaining silent (i.e., $\mathrm{a}_n=s$).
\end{itemize}

It is straightforward to see that the payoff of each player does not depend only on his own action but on the entire action profile since SINR function depends on both the chosen SU (i.e., the channel gain of the established link) and the activated SBSs (i.e., the amount of interference caused to the chosen SUs). Furthermore, we have assumed that one SU cannot be associated to more than one SBS, and hence two players choosing the same action in $\mathcal{M}$ has to be penalized by receiving a negative payoff in order to prevent the collision. The SINR expression is given in \eqref{sinr:a} where $\mathcal{S}_n$ is now equal to $\mathcal{S}_n'=\{n'\in\mathcal{N}: n'\neq n\;\text{and}\;\mathrm{a}_{n'}\neq s\}$.

The payoff function for each player $n\in\mathcal{N}$ is given explicitly by:
\begin{align}
\label{cases:2}
u_n(\mathbf{a}) = u_n(\mathrm{a}_{n}, \mathbf{a_{-n}})\myeq
\begin{cases}
0 &\mbox{if }\, \mathrm{C_1}\\
-1&\mbox{if }\, \mathrm{C_2}\\
1&\mbox{if }\, \mathrm{C_3}
\end{cases},
\end{align}
The conditions $\mathrm{C_1}$, $\mathrm{C_2}$, and $\mathrm{C_3}$ are given below:

\begin{itemize}
\item  $\mathrm{C_1}$: $\mathrm{a}_n = s$, i.e., the SBS $n$ chooses to remain silent;
\item  $\mathrm{C_2}$: $(\mathrm{a}_n = m$ and $\mathrm{SINR}_{\mathrm{a}_n n}\left(\mathrm{a}_n, \mathbf{a}_{-n}\right)\lt \beta)$ or $(\exists\,n'\neq n, \mathrm{a}_{n'}=m)$, i.e., the SBS $n$ chooses to transmit to SU $m$ and the required SINR is not met or another SBS $n'$ is choosing the same SU $m$; and
\item  $\mathrm{C_3}$: $(\mathrm{a}_n = m$ and $\mathrm{SINR}_{\mathrm{a}_n n}\left(\mathrm{a}_n, \mathbf{a}_{-n}\right)\geqslant \beta$ and $\forall\,n'\neq n, \mathrm{a}_{n'}\neq m)$, i.e., the SBS $n$ chooses to transmit to SU $m$ and the required SINR is met and no other SBS $n'\neq n$ is choosing SU $m$.
\end{itemize}

In game $\mathfrak{G}$, a PNE is a feasible solution to the optimization problem \eqref{unw:pb} that is stable, i.e., a subset of SBSs $\subseteq\mathcal{N}$ are transmitting to a subset of SUs $\subseteq\mathcal{M}$ and no single SBS $n\in\mathcal{N}$ has the incentive to deviate from this solution because their SINR is satisfied (the other SBSs are silent). Note that this is not the case for $\mathfrak{G}_1$ and $\mathfrak{G}_2$ which have PNEs that is not necessarily a feasible solution for the optimization problem \eqref{unw:pb} because of the collision. Unfortunately, the proposed game $\mathfrak{G}$ does not have PNEs, in general, as we show below.

\begin{proposition}
\label{propo:3}
    The game $\mathfrak{G}$ does not admit a PNE in general.
\end{proposition}

\begin{IEEEproof}
    We prove Proposition~\ref{propo:3} by constructing a counterexample, an instance of the game that does not admit a PNE. Let $N=M=3$, $\mathcal{N}=\{1,2,3\}$, $\mathcal{M}=\{1,2,3\}$, $p_n=4$ for all $n\in\{1, 2, 3\}$ and $\beta=2$. Further, assume that each SBS $i$ can only transmit to SU $i$ for all $i\in\{1,2,3\}$, SBS $1$ can transmit along with SBS $3$ while SBS $3$ cannot, SBS $2$ can transmit along with SBS $1$ while SBS $1$ cannot and SBS $3$ can transmit along with SBS $2$ while SBS $2$ cannot.
    Without loss of generality, the above configuration can be transformed to linear system of inequalities and one solution can be given by the following matrix:
	\begin{align}
		\mathbf{H}=
		\begin{bmatrix}
		1 & \frac{1}{4} & \frac{3}{10} \\[.3em]
		\frac{3}{10} & 1 & \frac{1}{4} \\[.3em]
		\frac{1}{4} & \frac{3}{10} & 1
		\end{bmatrix},
	\end{align}
	where $\mathbf{H} = \left[\mathrm{h}_{mn}\right]$ is the channel coefficients given by $\mathrm{h}_{mn}=|\mathrm{g}_{mn}|^2$.
	
	We claim that this game has no PNE which can be verified by testing all the possible action profiles using a brute force technique, for example. Hence there are instances of $\mathfrak{G}$ where no PNE exists. Therefore, $\mathfrak{G}$ does not always admit PNEs.
\end{IEEEproof}

Even though, the game $\mathfrak{G}$ is shown to have no PNEs in general, we observe through simulations (see Section~\ref{simu}) that it does admit PNE almost always. This observation is illustrated by the simulation of different channel realizations where all of them has PNEs. In other words, we claim that there exist always PNEs given the channel gain $|\mathrm{g}_{mn}|^2$ with the path loss and the Rayleigh fading.

\section{Efficiency of the Proposed Game and Discussions\label{disc}}

This section discusses the efficiency of the proposed game $\mathfrak{G}$. It also discusses the performance of the games $\mathfrak{G}_1$ and $\mathfrak{G}_2$ and compares the convergence of the BRD and the mWSLS algorithm to PNEs.

The efficiency of PNE can be measured by the PoA and the PoS~\cite{Koutsoupias, Agussurja}. We formally define these two ratios as follows:
\begin{definition}
	In the non-cooperative game $\mathfrak{G}$, denote the set of PNE action profiles by $\mathcal{P}^*\subseteq\mathcal{A}$. Then, the PoA and the PoS are the respective ratios: 
	\begin{align}
		\mathrm{PoA} = \dfrac{\min\limits_{\mathbf{a}\in\mathcal{P}^*}\sum\limits_{n=1}^{N}u_n(\mathbf{a})}{\max\limits_{\mathbf{a}\in\mathcal{A}}\sum\limits_{n=1}^{N}u_n(\mathbf{a})},
	\end{align}
	\begin{align}
		\mathrm{PoS} = \dfrac{\max\limits_{\mathbf{a}\in\mathcal{P}^*}\sum\limits_{n=1}^{N}u_n(\mathbf{a})}{\max\limits_{\mathbf{a}\in\mathcal{A}}\sum\limits_{n=1}^{N}u_n(\mathbf{a})}.
	\end{align}
\end{definition}

Note that the discrete nature of the action set of the players and the random nature of the wireless channel make the analytic calculation in closed-form of the PoA and the PoS a very difficult task. For this reason, we evaluate these two ratios using computer-based simulations. We used Gambit tools for game theory~\cite{gambit} integrated with Python. 

Despite proposition~\ref{propo:3}, simulation results show that the game $\mathfrak{G}$ admits PNEs almost always.
\begin{figure}[!t]
\centering
\includegraphics[height=8.9cm,keepaspectratio]{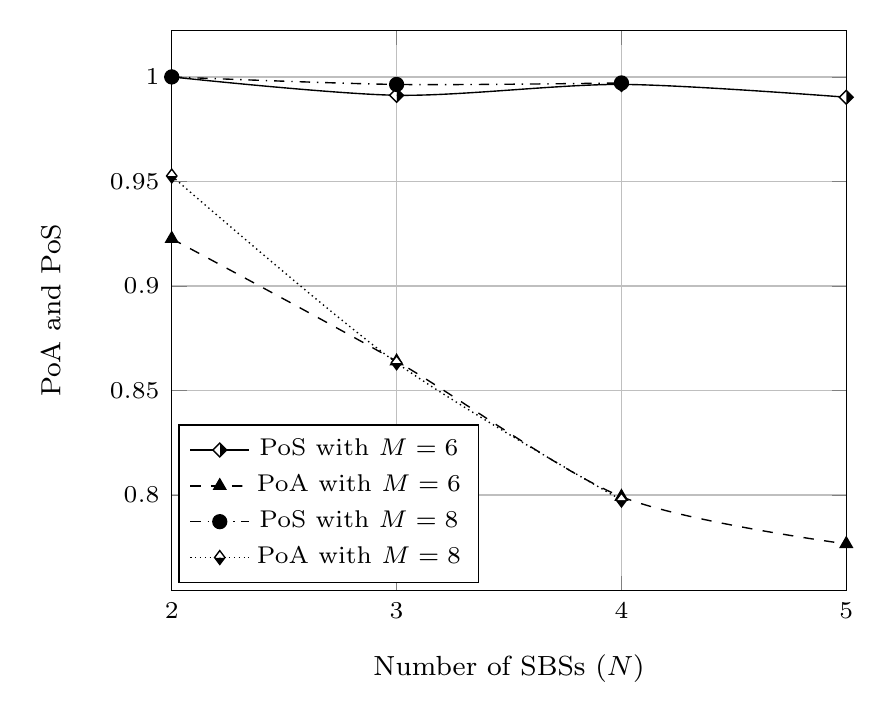}
\caption{PoA and PoS as a function of the number of players $N$ for the case of $6$ SUs and $8$ SUs.}
\label{fig:poa:pos}
\end{figure}
It is clear from Fig.~\ref{fig:poa:pos} that the proposed game $\mathfrak{G}$ is efficient especially when the number of players is small. That is, the PoA and the PoS are close to $1$. However, the PoA deteriorates as the number of players gets larger. This loss in the PoA means that the set of worst PNEs of the game are, in average, $23\%$ far from the social optimum when $N=5$. Even though the game $\mathfrak{G}$ has small PoA as the number of player gets larger, the PoA is still considered good as it is approximately equal to $77\%$ when $N=5$ and $M=8$. Fortunately, the PoS does not suffer from such a loss as depicted in Fig.~\ref{fig:poa:pos}. In fact, the figure shows that the PoS of the proposed game $\mathfrak{G}$ is almost equal to $1$ whatever the value of player $N$ is. This illustrates that the set of best PNEs of the game $\mathfrak{G}$ is almost equal to the set of social optimum. Therefore, if the PNE is well selected, then it must be a social optimum.

\begin{figure}
    \centering
    \includegraphics[height=8.9cm,keepaspectratio]{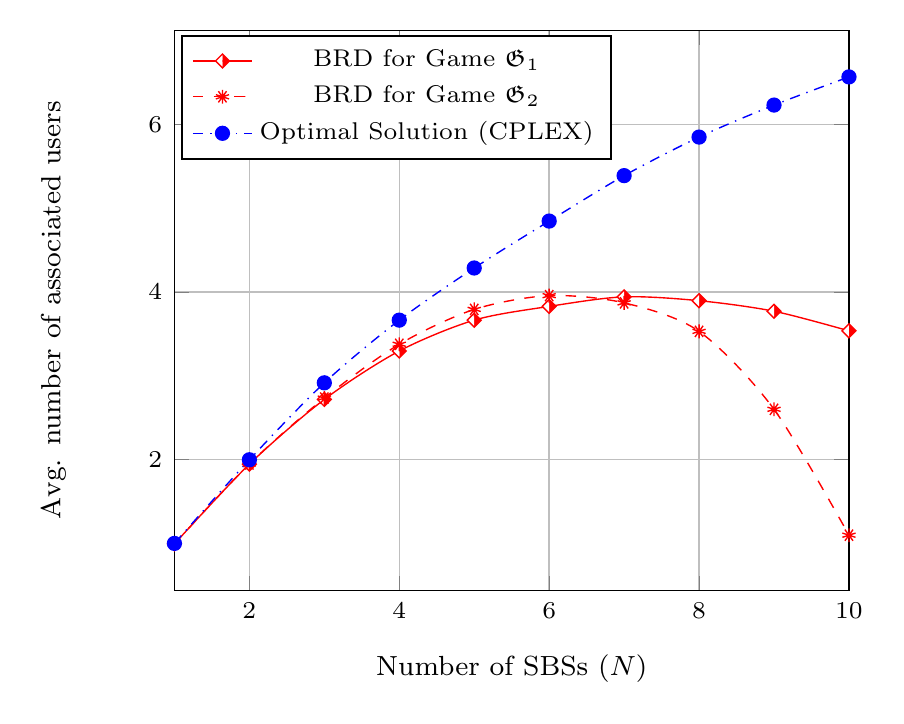}
    \caption{performance of the BRD for the game $\mathfrak{G}_1$ and $\mathfrak{G}_2$. $M=10$ }
    \label{fig:brd:opt:g1:g2}
\end{figure}

Next, we compare the performance of games $\mathfrak{G}_1$ and $\mathfrak{G}_2$ to the social optimum. To that end, we implemented the BRD algorithm for both games. In game $\mathfrak{G}_1$, when the BRD algorithm converges, the SBSs which have a payoff of $-1$ should not transmit whereas, in game $\mathfrak{G}_2$, the SBSs which have a payoff of $-1$ or a payoff of $-2$ (depending on the number of SBSs and SUs) should not transmit. The comparison between the social optimum, obtained by the CPLEX solver for problem \eqref{unw:pb}, and the PNEs of $\mathfrak{G}_1$ and $\mathfrak{G}_2$ given in Fig.~\ref{fig:brd:opt:g1:g2}.
This figure illustrates the bad performance of games $\mathfrak{G}_1$ and $\mathfrak{G}_2$. Despite the existence of PNEs in both games, these PNEs are far from the social optimum especially when $N$ gets larger. We also notice that the BRD algorithm for game $\mathfrak{G}_2$ gives better performance compared to the BRD algorithm for game $\mathfrak{G}_1$ when $N$ is small, i.e., $M>N$. This is due to the fact that the SBSs have more options to choose from and further there is no collision as pointed in the proof of proposition~\ref{prop:2}. When $N$ gets close to $M$ the BRD algorithm for game $\mathfrak{G}_2$ performs very badly and becomes clearly outperformed by the BRD algorithm for game $\mathfrak{G}_1$. 

Next, we propose two strategy updating algorithms that give good performance compared to the social optimum. The first algorithm is the BRD algorithm and the second one is inspired by the learning rule \textit{win-stay-lose-shift}~\cite{Nowak}.

\section{The Best Response Dynamics (BRD)\label{BRD}}

In order to solve the game $\mathfrak{G}$, we propose to use the BRD algorithm. Since the game $\mathfrak{G}$ does not admit PNEs, the BRD algorithm is not guaranteed, mathematically, to converge to a stable solution. The BRD algorithm starts out with a random user-BS association (random action profile, $\mathbf{a}$), obtained by the function \textsc{first$\_$round} in Algorithm~\ref{algo:BRD}. This randomness is obtained by the probability distribution $\bm\pi_{B}$ (see Algorithm~\ref{algo:BRD}). Then, BRD iterates either $\mathsf{R}$ times, where $\mathsf{R}$ is a maximum number of iterations, or until it converges (i.e., finds a stable solution), whichever occurs first. The action profile of players is determined in each iteration by calculating the set of best responses of each player. In each iteration, player $n$ updates his action based on his previously calculated best response set. Then, player $n'$ best responds accordingly and so on until all players updates their actions. The pseudo code of BRD is given in Algorithm~\ref{algo:BRD}.
\begin{algorithm}
  \caption{Best Response Dynamics For Game $\mathfrak{G}$ \label{algo:BRD}}
  \begin{algorithmic}[1]
    \Require{$\mathrm{g}_{nm}$, $M$, $N$, $p_n$, $\beta$}
    \Ensure{A near optimal solution $\mathbf{a^*}$}
    \Statex
    \Function{first\_round}{$\bm\pi_{B}$, $\mathcal{N}$}
        \LineComment{$\bm\pi_B$: A Probability distribution}
        \For{$n$ in $\mathcal{N}$}
	        \State SBS $n$ plays a random action $\mathrm{a}_n$ according to $\bm\pi_{B}$
  	    \EndFor
  	    \For{$n$ in $\mathcal{N}$}
	        \State SBS $n$ calculates its payoff $u_n(\mathrm{a}_n,\mathbf{a}_{-n})$
  	    \EndFor
  	    \State \Return{$\mathbf{a}=\left(\mathrm{a}_1,\cdots,\mathrm{a}_N\right)$, $\mathbf{u}=\left(u_1,\cdots,u_N\right)$}
    \EndFunction
    %\For{$n$ in $\mathcal{N}$}
    %    \State SBS $n$ calculates the payoff $u_n(\mathrm{a}_n, \cdot)$.
    %\EndFor
    %\State $\mathbf{\hat{a}}\gets \mathbf{a}$
    %\For{$n$ in $\mathcal{N}$}
    %    \If{$\hat{u}_n(\mathrm{a}_n, \cdot) == -1$ or $\hat{u}_n(\mathrm{a}_n, \cdot) == 0$}
    %        \State $\mathcal{S}\gets\mathcal{S}\cup\{n\}$
    %    \EndIf
    %\EndFor
%    \LineComment{$\mathcal{S}$: the players who have a payoff of $0$ or $-1$}
    \State $\mathsf{r},\mathsf{R} \gets 0,10$
    \LineComment{$\mathsf{R}$ is a variable used to guarantee the termination of the while loop}
    \While{$\left(\mathsf{r}<\mathsf{R}\right)$ or $\left(\mathbf{a}\,\text{ is not a}\, \mathsf{PNE}\right)$}
        \State $\mathcal{BR}\gets\varnothing$
		\For{$n$ in $\mathcal{N}$}
		    \For{$m$ in $\mathcal{A}_n$}
		        \If{$u_n(m, \mathbf{a}_{-n})>u_n(\mathrm{a}_n, \mathbf{a}_{-n})$}
		            \State $\mathcal{BR}\gets\{m\}$
		            \State $\mathrm{a}_n\gets m$
	            \ElsIf{$u_n(m, \mathbf{a}_{-n})==u_n(\mathrm{a}_n, \mathbf{a}_{-n})$}
		            \State $\mathcal{BR}\gets\mathcal{BR}\cup\{m\}$
	            \EndIf
		    \EndFor
      	    \State SBS $n$ selects an action $m$ from $\mathcal{BR}$\label{line:tie}
      	    \LineComment{Tie-breaking at random}
      	    \State $\mathrm{a}_n\gets m$
      	\EndFor
      	%\State Update $\mathbf{\hat{a}}$, $\mathbf{\hat{u}}$, and $\mathcal{S}$
      	\State $\mathsf{r}\gets \mathsf{r}+1$
      \EndWhile
      \State $\mathbf{a^*}\gets\mathbf{a}$
      %\State $ \mathbf{u^*}\gets\mathbf{\hat{u}}$
      \State \Return{$\mathbf{a^*}$}
  \end{algorithmic}
\end{algorithm}

The BRD algorithm may converge to a bad PNE due to 1) the choice made by the \textsc{first$\_$round} function; or 2) the tie-breaking rule given in line 22 of Algorithm~\ref{algo:BRD}. In Fig.~\ref{bad:poa:pos}, an instance of the game $\mathfrak{G}$ is given where the PoS $=1$ and the PoA $=1/2$. However, as shown in Fig.~\ref{bad:poa:pos}, the BRD algorithm converges to a PNE where $50\%$ of the SUs are associated ($\mathsf{u}2$ is associated and $\mathsf{u}1$ is not) even though the game $\mathfrak{G}$ admits a PNE where $100\%$ of the SUs are associated (both $\mathsf{u}2$ and $\mathsf{u}1$ are associated). The BRD algorithm reaches this bad PNE because it starts with the action profile $(s, \mathsf{u}1)$ as shown by the arrows in Fig.~\ref{bad:poa:pos}. If it had chosen another action profile, it would have reached the social optimum. In Fig.~\ref{bad:tie}, the BRD algorithm has to choose between two best responses ($(\mathsf{u}1, s)$ and $(\mathsf{u}2, s)$) starting from the action profile $(s, s)$ as shown by the arrows. If the BRD algorithm chooses $(\mathsf{u}2, s)$, it will converge to a PNE that is $50\%$ far from the social optimum whereas if it chooses $(\mathsf{u}1, s)$, it will converge to a PNE that is equal to the social optimum. 
In order to guarantee a tight-to-optimal solution of the BRD algorithm, it will be executed $Q$ times with the same probability distribution. By re-executing the BRD algorithm $Q$ times, it is guaranteed that the the best PNE will be selected more often (this will be shown in Section~\ref{simu}).
\begin{figure}[!h]
\centering
\begin{subfigure}{.5\textwidth}
  \centering
      
    \begin{tikzpicture}
    % the matrix entries
    \matrix (mat) [table]
    {
    & $\mathsf{u}1$ & $\mathsf{u}2$ & $s$ \\
    $\mathsf{u}1$ & (-1, -1)  & (1, 1) & (1, 0) \\
    $\mathsf{u}2$ & (1, -1) & (-1, -1) & (1, 0) \\
    $s$ & (0,1) & (0, 1) & (0,0) \\
    };
    % the matrix rules
    \foreach \x in {1,...,3}
    {
      \draw 
        ([xshift=-.5\pgflinewidth]mat-\x-1.south west) --   
        ([xshift=-.5\pgflinewidth]mat-\x-4.south east);
      }
    \foreach \x in {1,...,3}
    {
      \draw 
        ([yshift=.5\pgflinewidth]mat-1-\x.north east) -- 
        ([yshift=.5\pgflinewidth]mat-4-\x.south east);
    }    
    % the arrows
    \begin{scope}[shorten >=7pt,shorten <= 7pt]
    \draw[->]  (mat-4-2.center) -- (mat-3-2.center);
    \draw[->]  ([shift={(8pt,8pt)}]mat-3-2.center) -- ([shift={(8pt,8pt)}]mat-3-4.center);
    %\draw[->]  (mat-4-3.center) -- (mat-2-3.center);
    \end{scope}
    \end{tikzpicture}
    \caption{Bad convergence of the BRD algorithm\\ due to the \textsc{first$\_$round} function.}
    \label{bad:poa:pos}
\end{subfigure}%
\begin{subfigure}{.5\textwidth}
  \centering
  \begin{tikzpicture}
    % the matrix entries
    \matrix (mat) [table]
    {
    & $\mathsf{u}1$ & $\mathsf{u}2$ & $s$ \\
    $\mathsf{u}1$ & (-1, -1)  & (1, 1) & (1, 0) \\
    $\mathsf{u}2$ & (1, -1) & (-1, -1) & (1, 0) \\
    $s$ & (0,1) & (0, 1) & (0,0) \\
    };
    % the matrix rules
    \foreach \x in {1,...,3}
    {
      \draw 
        ([xshift=-.5\pgflinewidth]mat-\x-1.south west) --   
        ([xshift=-.5\pgflinewidth]mat-\x-4.south east);
      }
    \foreach \x in {1,...,3}
    {
      \draw 
        ([yshift=.5\pgflinewidth]mat-1-\x.north east) -- 
        ([yshift=.5\pgflinewidth]mat-4-\x.south east);
    }    
    % the arrows
    \begin{scope}[shorten >=7pt,shorten <= 7pt]
    \draw[->]  (mat-4-4.center) -- (mat-3-4.center);
    \draw[->]  ([shift={(2.5pt,2.5pt)}]mat-4-4.center) -- ([shift={(2.5pt,2.5pt)}]mat-2-4.center);
    \draw[->]  ([shift={(8pt,8pt)}]mat-2-4.center) -- ([shift={(8pt,8pt)}]mat-2-3.center);
    \end{scope}
    \end{tikzpicture}
    \caption{Bad convergence of the BRD algorithm\\ due to the tie-breaking rule.}
    \label{bad:tie}
\end{subfigure}
\caption{Instance of $\mathfrak{G}$ with different convergence of the BRD algorithm.}
    \label{bad:brd}
\end{figure}
In this way, the probability that the BRD algorithm converges to a social optimum will approach one. However, this solution suffers from high information exchange and is computationally complex. Therefore, 
%Indeed, as $Q$ increases, the computational complexity increases too. Note also that line 11 of Algorithm~\ref{algo:BRD} guarantees the convergence of the BRD algorithm to a stable solution but it is a costly step because it has to verify if all the SBSs can deviate or not. Also, in line 15 of Algorithm~\ref{algo:BRD}, in order for SBS $n$ to best respond, it has to know the action of other SBSs $\mathbf{a}_{-n}$. This illustrates the huge complexity of such algorithm.
in the next section, we propose a completely distributed algorithm that has a better trade-off between computation, information exchange and performance.

\section{The Modified Win-Stay-Lose-Shift Algorithm (mWSLS)\label{mwsls}}

In this section we describe a completely distributed algorithm. It is called mWSLS for \textit{modified win-stay-lose-shift}. This algorithm is inspired by the well-known win-stay-lose-shift learning algorithm~\cite{Nowak}. We first describe the first iteration of the algorithm and its parameters in a pseudo-code and then describe the learning process.

\subsection{Overview and the First Iteration}

Each SBS $n$ runs the same learning algorithm. It starts by playing randomly an action $\mathrm{a}_n$ from its action space $\mathcal{A}_n$. Then, it starts transmitting to the chosen SU $\mathrm{a}_n=m$. The SU $m$ then computes its SINR and feeds back to the SBS $n$ whether the computed SINR is above the threshold or not. Specifically, the SU $m$ sends one bit of data which allows the SBS $n$ to have an information about how good its choice is. Clearly, this feedback depends not only on the action of the SBS $n$ but on the entire action profile of all the SBSs played during this first iteration. Hence, the SBS $n$ computes its reward based on the feedback as follows:

\begin{align}
\label{cases:5}
    \mathrm{u}_n(\mathrm{a}_{n}, \mathbf{a}_{-n})=
    \begin{cases}
        0 &\mbox{if }\, \mathrm{a}_n = s\\
        -1&\mbox{if }\, \mathrm{b} = 0\\
        1&\mbox{if }\, \mathrm{b} = 1,
    \end{cases}
\end{align}
where $\mathrm{b}$ is the feedback of the chosen SU $m=\mathrm{a}_n$. Note that to prevent the colliding state, where two or more SBSs choose the same SU, each SBS broadcasts its chosen action in the network.

Based on the received feedback and the computed rewards, the SBS that chooses to transmit during the current slot have an information about the quality of the chosen action. Whereas the SBS which chooses to remain silent during that slot has no new information. Hence, every time a SBS learns something new about the system, it must exploit it in order to play a better action in the future. Therefore, we associate to each SBS $n\in\mathcal{N}$, a probability vector $\bm\pi_n = [\pi_n[1], \dotsc, \pi_n[M], \pi_n[s]]^T$ of size $M+1$. Each element $\pi_n[m]$ corresponds to the probability of playing action $\mathrm{a}_n=m$ by SBS $n$. For the first iteration, we assume that the probabilities in each vector $\bm\pi_n$ are uniformly distributed, i.e., $\pi_n[m] = \frac{1}{M+1}, \forall m \in \{1,\dotsc,M\}\cup\{s\}$. Note that each SBS picks its first action according to this distribution. 

\subsection{The Learning Process}

Once the SBSs acquire the feedback of the corresponding SUs and map them into rewards, they proceed by updating their probability vectors. These updates are the core element of the learning process. In the original WSLS algorithm, each player starts by playing a random action. If the played action results in a higher payoff, then it is considered as a winning action and the player keeps playing it during the next round. Otherwise, this action is considered as a loosing one and the player has to shift into another action with the hope of improving its new reward. The WSLS is widely and efficiently applied when the rewards are Boolean (either $0$ or $1$). However, such learning strategy must be adapted when the rewards are finite but not Boolean which is the case in this paper where the rewards are in $\{-1,0,1\}$.

When the reward is equal to $1$, the action played in the current iteration is considered as winning action. Hence, the probability of playing it must be increased in order to augment the chances to converge to this action at the end of the learning process. Precisely, the probability corresponding to the winning action is updated as follows

\begin{equation}
	\pi^{t+1}_n[m] = \pi^t_n[m] + \tau \left(1-\pi^t_n[m]\right),
\end{equation}
where $m$ is the index of the winning action, $t$ is the iteration index\footnote{The iteration index is only used when necessary} and $\tau$ represents the \emph{winning increment factor} by which the probability of choosing the winning action during the next iteration is augmented. Note that all the probabilities in $\bm\pi_n$ other than $\pi_n[m]$ should be reduced by the same factor in order to keep their sum (including $\pi_n[m]$) equal to one. Hence, these probabilities are updated as follows:

\begin{equation}
	\pi^{t+1}_n[m'] = \pi^t_n[m'] - \tau \pi^t_n[m'], \quad \forall m' \neq m.
\end{equation}

When the reward is equal to $-1$, the action chosen in the current iteration is a loosing one. The probability of playing such action in the next iteration is reduced and the probability of remaining silent (i.e., playing a action $s$) is increased. This learning strategy is motivated by the fact that when a SBS plays many loosing actions, it is then preferable to force it to learn playing the silence action. The probability vector is therefore updated as follows:

\begin{equation}
	\pi^{t+1}_n[m] = \pi^t_n[m] - \varepsilon,
\end{equation}
\begin{equation}
	\pi^{t+1}_n[s] = \pi^t_n[s] + \varepsilon,
\end{equation}
where $m$ is the index of the loosing action, $s$ is the index of the silent action and $\varepsilon$ represents the \emph{loosing decrement factor}. When the algorithm performs the mentioned updates, it should always check that the probabilities remain consistent, i.e., ($0 < \pi_n[m] < 1,~\forall n,m$) and ($\sum_m \pi_n[m] = 1,~\forall n$).

\begin{algorithm}
    \caption{Pseudo-code of the part of the mWSLS algorithm executed in each SBS\label{algo:mWSLS}}
    \begin{algorithmic}[1]
        \Require{$\mathcal{A}_n$, $\varepsilon$, $\tau$, $\bm\pi_n$}
        \Ensure{An action $\mathrm{a}_n$}
        \Statex
        \LineComment{$\bm\pi_n$ is a probability vector given by $\bm\pi_n= [\pi[1,n], \dotsc, \pi[M,n], \pi[s,n]]^T$}
        \If{$u_n(\mathrm{a}_n, \cdot) == 1$}
            \State $\pi[\mathrm{a}_n, n]\gets \tau+\pi[\mathrm{a}_n, n]\cdot(1-\tau)$
            \State $\mathsf{tmp}\gets\mathcal{A}_n\backslash\{\mathrm{a}_n\}$
            \State $\pi[\mathsf{tmp}, n] = \pi[\mathsf{tmp}, n]\cdot(1-\tau)$
        \ElsIf{$u_n(\mathrm{a}_n, \cdot) == -1$}
            \State $\pi[\mathrm{a}_n, n]\gets\pi[\mathrm{a}_n, n]-\varepsilon$
            \State $\pi[s, n] = \pi[s, n] + \varepsilon$
        \EndIf
        % \State Player $n$ finds his set of best responses
        % \State Select one best response $\hat{\mathrm{a}_n}$ \label{line:tie}
        % \LineComment{Tie-breaking at random}
        % \State SBS $n$ plays an action $\mathrm{a}_n$ according to $\bm\pi_n$
        %\State SBS $n$ calculates the payoff $u_n\left(\mathrm{a}_n, \cdot\right)$
        \State \Return{$\mathrm{a}_n$}
    \end{algorithmic}
\end{algorithm}

After performing a given number of iterations, each SBS learns to play either a winning action which allows it to obtain a positive reward or the silence action. All SBSs that are chosen to transmit converge to a steady state where the probability vectors contain a single one positioned in the winning action corresponding to the associated SU. Whereas the SBSs which converge with a probability vector containing one in the index $s$ has to remain silent during the current time-slot. The pseudo code of the mWSLS algorithm executed in each SBS is given by Algorithm~\ref{algo:mWSLS}.

\section{Simulation Results\label{simu}}

In this section, the performance of the proposed learning algorithm mWSLS and the BRD algorithm obtained by simulations are presented. We first investigate the impact of the different parameters of both algorithms on their performance. Then, we present the efficiency of the mWSLS and the BRD and compare them against the social optimum solution in terms of average number of associated users.

For all the simulations, the coefficient $\mathrm{g}_{mn}$ is given mathematically by $\rho_{mn}\sqrt{(\frac{d_0}{d_{mn}})^\alpha}$, where $\alpha$ is the path loss coefficient, $d_{mn}$ is the distance between $m$ and $n$, uniformaly generated, $d_0$ is a reference distance at which the  path loss is calculated and $\rho_{mn}$ is the long-term fading modeled as a zero-mean, complex Gaussian random variable with unit variance. Unless otherwise specified, the SINR threshold required by each SU is set to $\beta = 0$ dB, the transmit power at each SBS $n$ is fixed to $p_n = 10$ dB (this power is normalized by the noise power and the reference distance $d_0$) and the path loss exponent is set to $\alpha=4$. The long-term fading effect is assumed to be Rayleigh fading. The number of iterations of the mWSLS is chosen to be $T=100$. The results of all the performed simulations are averaged over $5000$ channel realization.

\begin{figure}[!h]
\centering
\includegraphics[height=8.9cm,keepaspectratio]{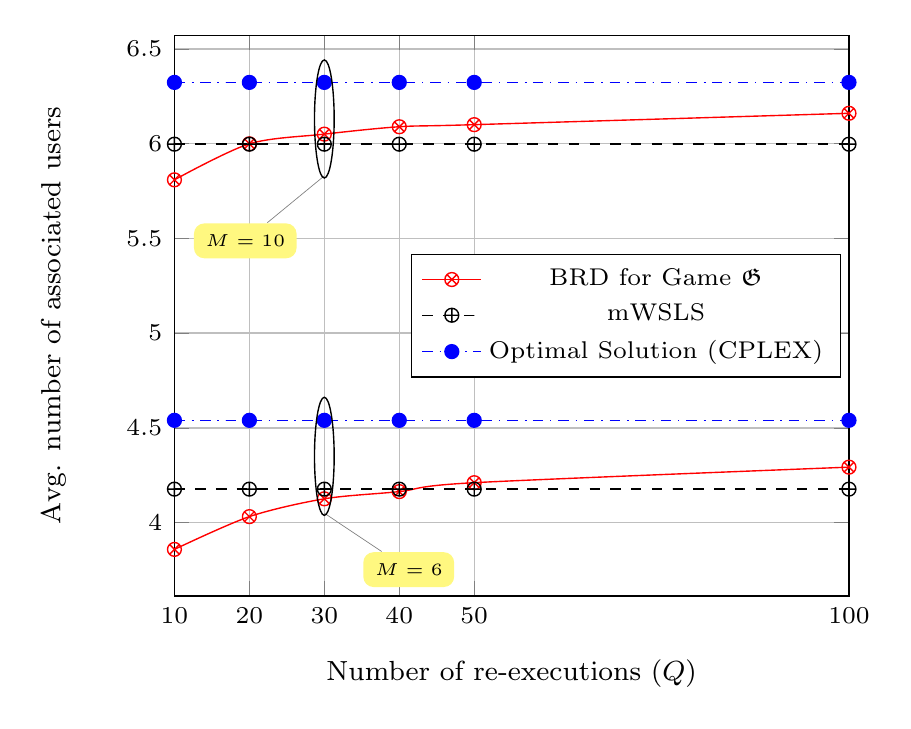}
\caption{Average number of associated SUs vs. the number of re-execution of  BRD algorithm $Q$}
\label{fig:brd:iter}
\end{figure}

\begin{figure}[!h]
\centering
\includegraphics[height=8.9cm,keepaspectratio]{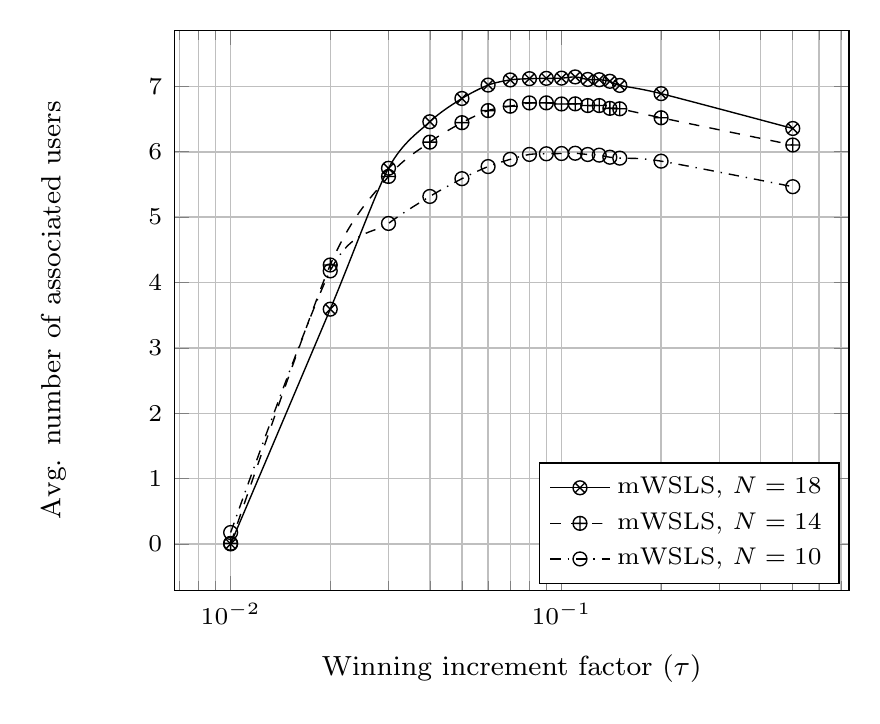}
\caption{Average number of associated SUs vs. $\tau$}
\label{fig_tau_N}
\end{figure}

On the one hand, the most important parameter of the BRD algorithm is the number of times it has been executed, $Q$. In Fig.~\ref{fig:brd:iter}, we compare, for a fixed number of SBSs $N=10$ and for a fixed number of SUs $M=6$ and $M=10$, the mWSLS and the social optimum against the BRD algorithm. The curve of the BRD algorithm in Fig.~\ref{fig:brd:iter} is obtained by executing the BRD algorithm $Q$ times where $\bm\pi_{B}$ is the uniform distribution.

%In Fig~\ref{fig:brd:iter}, the performance of the BRD algorithm, which is the average number of associated users, is plotted as a function of the number of iterations, $\mathsf{R}$. 
Fig.~\ref{fig:brd:iter} shows that when $Q$ is small, the BRD algorithm performs worst than the mWSLS algorithm and is far from the social optimum. This is due to the bad first association given by the first round of the algorithm as explained in Fig.~\ref{bad:poa:pos}. On the other hand, when $Q$ increases, the performance of the BRD algorithm improves and outperforms the proposed mWSLS algorithm. This is essentially because the BRD algorithm would explore further the set of PNEs and hence, with high probability, the BRD algorithm approaches the best PNE with high probability. We mention that when the action space is large, the BRD algorithm outperforms the mWSLS algorithm with less $Q$. (Less than $30$ re-executions are needed for $M=10$ compared to $50$ re-executions for $M=6$.)

\begin{figure}[!h]
\centering
\includegraphics[height=8.9cm,keepaspectratio]{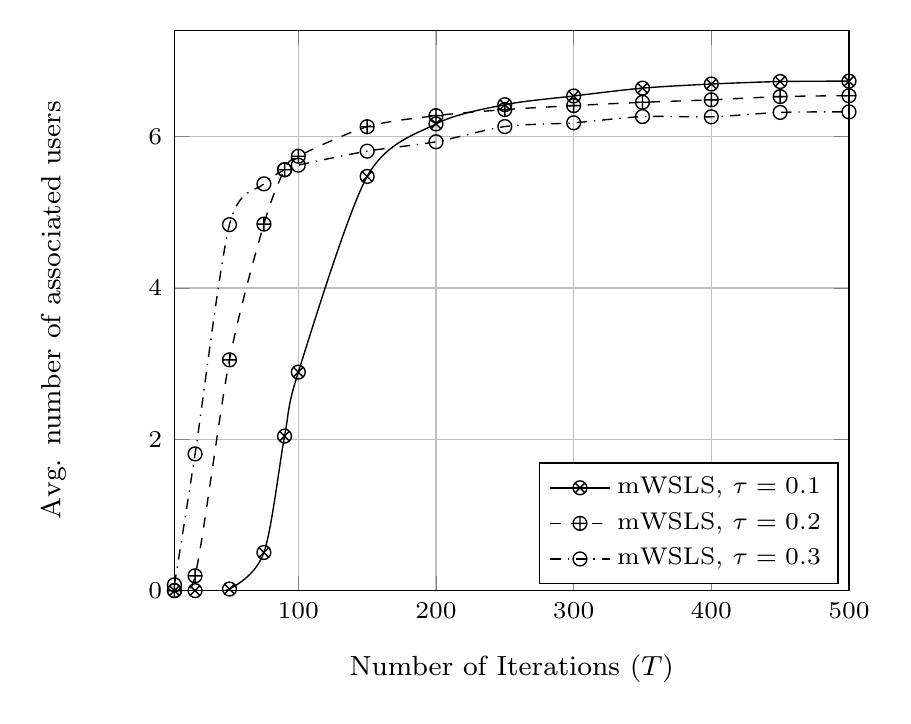}
\caption{Average number of associated SUs vs. $T$}
\label{fig_iter}
\end{figure}

\begin{figure}[!h]
\centering
\includegraphics[height=8.9cm,keepaspectratio]{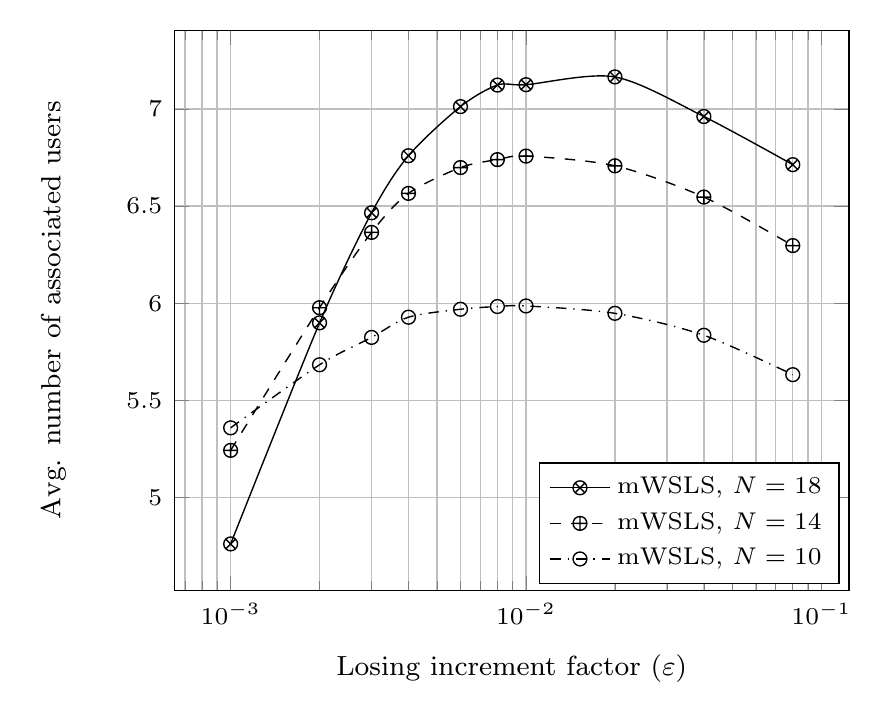}
\caption{Average number of associated SUs vs. $\varepsilon$}
\label{fig_epsilon_N}
\end{figure}

The performance and the convergence of the mWSLS algorithm rely on the choice of three parameters, namely the \textit{winning increment factor} $\tau$, the \textit{loosing decrement factor} $\varepsilon$ and the number of iterations $T$. Hence, we discuss here their impact on the performance. Fig.~\ref{fig_tau_N} plots the average number of associated SUs as a function of the factor $\tau$. We notice that the impact of $\tau$ on the performance increases as the number of SBSs increases. Also, it can be observed that the optimal value for the three scenarios is around $0.1$. However, we can notice that the value of optimal $\tau$ tends to get smaller as $N$ gets larger. The number of associated SUs as a function of the iteration index for different values of $\tau$ are shown in Fig.~\ref{fig_iter}. We notice that the optimal $\tau$ results in higher performance but suffers from very slow convergence. Whereas a large $\tau$ achieves poor performance but converges very quickly, e.g., $\tau = 0.3$ achieves a performance of serving three SUs in less than $15$ iterations while the same performance is obtained after more than a hundred iterations with $\tau = 0.1$. Hence, the factor $\tau$ can be seen as a tuning parameter to adjust the performance/convergence time trade-off. Fig.~\ref{fig_epsilon_N} plots the algorithm performance as a function of the factor $\varepsilon$ for different numbers of SBSs $T$. The figure shows that the optimal value of $\varepsilon$ is around $0.01$ and has a noticed impact on the performance of the learning. This impact becomes important when $N$ gets larger. This is because when there are more SBSs in the network, the number of silent SBSs is more important.

\begin{figure}[!h]
\centering
\includegraphics[height=8.9cm,keepaspectratio]{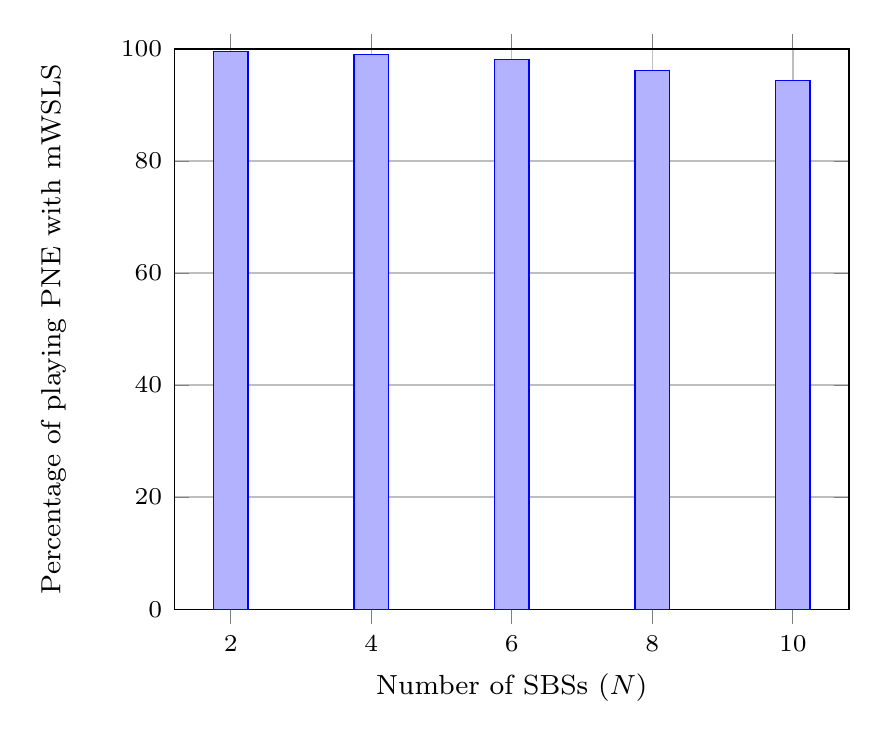}
\caption{Percentage of playing a PNE in the game $\mathfrak{G}$ with mWSLS.}
\label{fig:nash}
\end{figure}

\begin{figure}[!h]
\centering
\includegraphics[height=8.9cm,keepaspectratio]{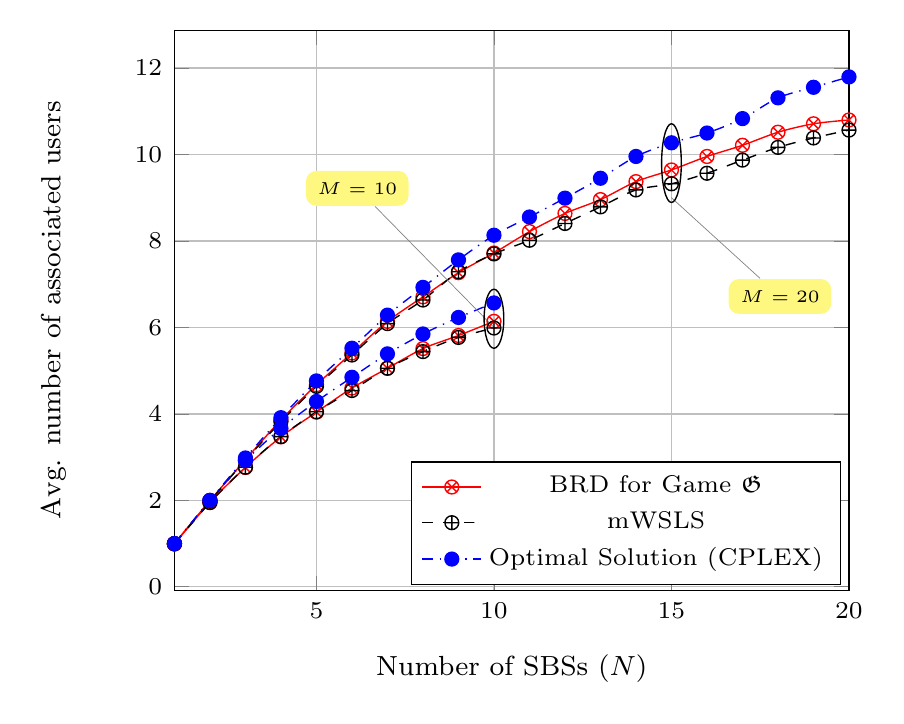}
\caption{Performance comparison of mWSLS, BRD with $Q=30$ and social optimum.}
\label{algo}
\end{figure}

Fig.~\ref{fig:nash} presents the percentage of time the mWSLS algorithm converges to a PNE. As stated before, there exist no known completely distributed algorithm that ensures to converge to such an equilibrium. However, Fig.~\ref{fig:nash} shows that the mWSLS algorithm allows the SBSs to play a PNE for a long period of time, especially for small $N$. Also, even when $N$ is equal to $10$ the algorithm converges to a PNE for more than $95\%$ of the time.

In Fig.~\ref{algo}, we present the performance of the mWSLS algorithm against the optimal solution, derived by the branch-and-bound algorithm (CPLEX), and the BRD for game $\mathfrak{G}$. Both BRD and mWSLS are close to the social optimum even for large values of $M$ and $N$. Further, the performance gap between the two strategy updating algorithm is still the same when $N$ gets larger. Hence, this small gap suggests the use of the completely distributed algorithm mWSLS in practice.
%\clearpage
\section{Conclusion\label{cl}}

This paper studies the user-BS association problem under quality of service requirements in HetSNets. Centralized and optimal algorithms suffer from many implementation issues. On the one hand, they need a large amount of feedback which relies on resource consuming information exchange between the SBSs and the SUs. Furthermore, they suffer from a huge computational complexity since the optimized problem is NP-hard. Thus, in this paper, we modeled this problem using non-cooperative game theory. First, we showed that the two first formulated games admit PNE but suffer from bad performance. Hence, we designed and studied a better game model. We showed that even though the game does not admit a PNE in general settings, but when it does then it has a PoA and PoS very close to $1$. Next, we proposed the BRD algorithm to solve the game. We also proposed a completely distributed algorithm which assumes no coordination between the SBSs. This algorithm is based on the win-stay-lose-shift learning model (called mWSLS). Through simulation, we assessed the performance of the proposed game and the proposed algorithms. Also, the mWSLS algorithm is shown to converge to a near optimal user association solution approaching the BRD algorithm and the complex centralized performance obtained by a computationally complex algorithm which assumes complete channel information. Furthermore, the mWSLS algorithm converges to PNE with high probability.
% \appendices

% \section{Proof of the First Zonklar Equation}

% \section*{Acknowledgment}
% \clearpage

\bibliographystyle{IEEEtran}
\bibliography{IEEEabrv,bibliotex}
%\balance

\end{document}